\documentstyle[seceq]{ptptex}
\newcommand{\overslash}[1]{\ooalign{\hfil/\hfil\crcr$#1$}}
\newcommand{\dslash}{\ooalign{\hfil/\hfil\crcr$\partial$}}
\newcommand{\Dslash}{\ooalign{\hfil/\hfil\crcr$D$}}

\newcommand{\mbold}[1]{\mbox{\boldmath{$#1$}}}
\begin{document}
\notypesetlogo
\markboth{J. Sakamoto and K. Mano}{Comments on Bosonization of the
extended Thirring Model }
\title{Comments on  Bosonization of the extended Thirring Model with $SU(2)$ Symmetry }
\author{Jiro {\sc Sakamoto}\footnote{E-mail:\ jsakamot@riko.shimane-u.ac.jp}\ and Keiji {\sc Mano}}
\inst{Department of Material Science, Shimane University, Matsue 690-8504, Japan}
\abst{
Bosonization of the extended Thirring model with $SU(2)$ symmetry in the
Minkowski path integral method is discussed. We argue that it is not an
easy task to bosonize such a model if we derive correctly the fermion
determinant which is induced with the decoupling transformation because
it seems that there arise ghost fields. This is
contrary to what is shown in a textbook and some people believe.

}
\maketitle
\section{Introduction}

In a previous paper \cite{J_M} we derived the direct correspondence
between the fermion currents and the boson fields for the extended
Thirring model with $SU(2)$ symmetry in the Minkowski path integral
method, and we showed that the corresponding boson model contains ghost
fields which interact with the other boson fields in a complex manner
and that it is hard to eliminate them, while in the original Thirring
model such ghost field also arises but it is decoupled from other fields
and disappears in the final Lagrangian.

Since Coleman carried out the bosonization of the massive Thirring
model and showed that it is equivalent to the sine-Gordon model,\cite{Coleman} many people have attempted to apply this technique
to fermion models with non-Abelian symmetries,\cite{review} and some people believe
that the bosonization of the Thirring model with $SU(N)$ symmetry can be
easily carried out.\cite{Justin} Their consequences, however, do not
agree with ours mentioned above. In the present paper we argue that
in their formulation the fermion determinant, which is induced by the
decoupling transformation of the fermion fields, is not correctly
derived.

In the next section, after reviewing the formulation of
Ref.~\citen{Justin} briefly we derive the correct expression of the
fermion determinant for the decoupling transformation following
Fujikawa's method.\cite{review,Fujikawa,J_P} We use the same notation as
in the previous paper.\cite{J_M}: $g_{\mu\nu}=(-1,+1)$,
$\epsilon^{01}=-\epsilon_{01}=1$, $\gamma^0=i\sigma_x$,
$\gamma^1=\sigma_y$ and $\gamma^5=\sigma_z$. The $2\times 2$ matrices of
the $SU(2)$ generators are given by $\tau_1=\frac{1}{2}\sigma_x,
\tau_2=\frac{1}{2}\sigma_y$ and $\tau_3=\frac{1}{2}\sigma_z$. We use the first part of the Roman alphabet
as $SU(2)$ suffixes 1, 2, 3 and the middle part $i,j$ as only 1 and 2.

\setcounter{equation}{0}
\section{Derivation of the fermion determinant}

Our initial Lagrangian for the extended Thirring model with $SU(2)$
symmetry is given by
\begin{equation}
 {\cal L}=\overline{\psi}\dslash\psi + \frac{g}{2}\sum_{a=1}^3
j_{\mu}^aj_a^{\mu} + \frac{g'}{2}j_\mu j^\mu,\label{ini_L}
\end{equation}
with the $SU(2)$ current 
\begin{equation}
 j_{\mu}^a=i\overline{\psi}\tau_a\gamma_{\mu}\psi
\end{equation}
and the $U(1)$ current
\begin{equation}
j_{\mu}=i\overline{\psi}\gamma_{\mu}\psi.
\end{equation}
The last term in Eq.(\ref{ini_L}) is necessary for renormalizability 
of the model. The generating functional is given by
\begin{equation}
 Z=\int d\overline{\psi}d\psi\exp i\int d^2x {\cal L}.
\end{equation} 

Now we review the formulation of Ref.~\citen{Justin} briefly. We divide
Eq.(\ref{ini_L}) into two parts as follows,
\begin{eqnarray}
 {\cal L}_0 &=& \overline{\psi}\dslash\psi + \frac{g}{2}j_{\mu}^3j_3^{\mu}
 + \frac{g'}{2}j_{\mu}j^{\mu},\label{L_0}\\
{\cal L}_I &=& \frac{g}{2}\sum_{i=1}^2 j_{\mu}^ij_i^{\mu}.\label{L_I}
\end{eqnarray}
${\cal L}_0$ of the above expressions consists of  the diagonal parts of the $SU(2)$ multiplets and is
treated as the non-perturbative part of the Lagrangian, and ${\cal L}_I$ is treated as the perturbative part. Eq.(\ref{L_0}) is equivalently rewritten as
\begin{equation}
 {\cal L}_0 = \overline{\psi}\dslash\psi + j_{\mu}^3 A_3^{\mu} -\frac{1}{2g}A_{\mu}^3A_3^{\mu} + j_{\mu}a^{\mu} - \frac{1}{2g'}a_{\mu}a^{\mu},\label{L_0_1}
\end{equation}
with auxiliary vector fields $A_{\mu}^3$ and $a_{\mu}$. 
In two-dimensional spacetime, these vector fields are expressed as
\begin{eqnarray}
 A_{\mu}^3 &=& \partial_{\mu}\Lambda_3 + \epsilon_{\mu\nu}\partial^{\nu}\Phi_3,\\
a_{\mu} &=& \partial_{\mu}\lambda + \epsilon_{\mu\nu}\partial^{\nu}\phi.
\end{eqnarray} 
We substitute the above expressions into Eq.(\ref{L_0_1}) to obtain
\begin{eqnarray}
 {\cal L}_0 &=& \overline{\psi}\{\dslash + i\tau_3(\dslash \Lambda_3 + \gamma_5\dslash\Phi_ 3)+i(\dslash \lambda + \gamma_5\dslash\phi)\}\psi \nonumber\\
&& - \frac{1}{2g}\{\partial^{\mu}\Lambda_3\partial_{\mu}\Lambda_3 - \partial^{\mu}\Phi_3\partial_{\mu}\Phi_3\} - \frac{1}{2g'}\{\partial^{\mu}\lambda\partial_{\mu}\lambda - \partial^{\mu}\phi\partial_{\mu}\phi\},\label{L_0_2} 
\end{eqnarray}
where we use the identity $ \gamma_5\gamma_{\mu}=
-\epsilon_{\mu\nu}\gamma^{\nu}$.
The generating
functional for ${\cal L}_0$ is given by 
\begin{equation}
 Z_0 = \int d\overline{\psi}d\psi d{\Lambda}_3 d\Phi_3d\lambda d\phi\exp i\int d^2x {\cal L}_0.
\end{equation}

The coupling terms between the fermion and boson fields in (\ref{L_0_2}) 
are eliminated by the following decoupling transformation,
\begin{subeqnarray}
&& \psi\rightarrow \psi ' = U(\Lambda_3, \Phi_3,\lambda,\phi)\psi,\label{decouple_trans}\\
&&U(\Lambda_3, \Phi_3, \lambda, \phi)= \exp i\{\tau_3(\Lambda_3-\gamma_5 \Phi_3)+ \lambda -\gamma_5\phi\},
\end{subeqnarray}
and we have
\begin{equation}
 {\cal L}_0 = \overline{\psi '}\dslash\psi' - \frac{1}{2g}\{\partial^{\mu}\Lambda_3\partial_{\mu}\Lambda_3 - \partial^{\mu}\Phi_3\partial_{\mu}\Phi_3\} - \frac{1}{2g'}\{\partial^{\mu}\lambda\partial_{\mu}\lambda - \partial^{\mu}\phi\partial_{\mu}\phi\}.
\end{equation}
 The kinetic terms of $\Phi_3$ and $\phi$ in the above expression have 
wrong signs and they seem to be ghost fields. As is seen in the
following, however, $\phi$ is decoupled from the other fields and
$\Phi_3$ is absorbed to the other fields and they finally disappear.

The fermion Jacobian which is induced with the transformation
eq.(\ref{decouple_trans}) is derived following Fujikawa's method as\cite{J_M,review,Fujikawa}
\begin{equation}
 J = \exp i\int d^2x \frac{1}{4\pi}\{(\partial\Phi_3)^2 + 2(\partial\phi)^2\},
\label{Jacobian1}\end{equation}
where we use the eigenfunction $\psi_n(x;t)$ defined by
\begin{equation}
 i\Dslash(t)\psi_n(x;t)\equiv i(\dslash + i \overslash{\mbold{a}}(t) )\psi_n(x;t)= \lambda_n\psi_n(x;t)\label{eigen_eq1}
\end{equation}
as the basis functions for the Jacobian. Here we put
\begin{eqnarray}
 \overslash{\mbold{a}}(t)&=& U(-\Lambda_3,\Phi_3,-\lambda,\phi;-t)\overslash{\mbold{a}}U(\Lambda_3,\Phi_3,\lambda,\phi;-t)\nonumber\\
&&-iU(-\Lambda_3,\Phi_3,-\lambda,\phi;-t)\dslash U(\Lambda_3,\Phi_3,\lambda,\phi;-t),
\end{eqnarray}
with
\begin{eqnarray}
 \overslash{\mbold{a}} =  A_3^\mu\gamma_\mu \tau_3+a^\mu\gamma_\mu = (\dslash\Lambda_3 + \gamma_5\dslash\Phi_3)\tau_3+\dslash\lambda + \gamma_5\dslash\phi,\\
U(\Lambda_3,\Phi_3,\lambda,\phi;t)=\exp it\{\tau_3(\Lambda_3-\gamma_5\Phi_3)+\lambda-\gamma_5\phi\}.
\end{eqnarray}
Eq.(\ref{eigen_eq1}) corresponds to the Euler equation for the fermion
field obtained from ${\cal L}_0$ of (\ref{L_0_1}),
\begin{equation}
 i(\dslash +i \overslash{\mbold{a}})\psi = 0.
\end{equation}
By the decoupling transformation (\ref{decouple_trans}), currents
$j_\mu^i$ ($i=1,2$) in ${\cal L}_I$ of (\ref{L_I}) are transformed as
\begin{equation}
 j_{\mu}^i= i \overline{\psi}\tau_i\gamma_\mu\psi = i\overline{\psi '}\tau_i\gamma_\mu\exp\{-2i\tau_3(\Lambda_3-\gamma_5\Phi_3)\}\psi ',
\end{equation}
and ${\cal L}_I$ is rewritten as
\begin{equation}
 {\cal L}_I = -\frac{g}{2}\overline{\psi'}_1\gamma_\mu\exp\{ -i\gamma_5\Phi_3\}\psi'_2\overline{\psi'}_2\gamma^{\mu}\exp \{i\gamma_5\Phi_3\}\psi'_1,\label{L_I_prim} 
\end{equation}
where we put $\psi'$ of $SU(2)$-doublet as
\begin{equation}
 \psi' = \left(\begin{array}{@{\,}c@{\,}}
\psi_1'\\
\psi_2'
\end{array}\right).
\end{equation}
Eq.(\ref{L_I_prim}) is rewritten as
\begin{eqnarray}
 {\cal L}_I &=& g\left\{\overline{\psi'}_{1+}\psi_{2+}'\overline{\psi'}_{2-}\psi_{1-}'\exp \{-2i\Phi_3\}+ \overline{\psi'}_{1-}\psi_{2-}'\overline{\psi'}_{2+}\psi_{1+}'\exp \{2i\Phi_3\}\right\}\nonumber\\
&=& -g\left\{\overline{\psi'}_{1+}\psi_{1-}'\overline{\psi'}_{2-}\psi_{2+}'\exp\{-2i\Phi_3\}+\overline{\psi'}_{1-}\psi_{1+}'\overline{\psi'}_{2+}\psi_{2-}'\exp\{2i\Phi_3\}\right\}\ \ \ \ \ \ \ 
\end{eqnarray}
with putting spin-doublet as
\begin{equation}
 \psi_{i}'= \left(\begin{array}{@{\,}c@{\,}}
\psi_{i+}'\\
\psi_{i-}'
\end{array}\right).
\end{equation}
We note that all the boson fields except $\Phi_3$ are decoupled from the
fermion fields and can be eliminated by the path integration. In a
perturbative expansion over ${\cal L}_I$ we see that we can replace the
fermion fields with their boson counterparts $\theta_i$ and obtain the effective Lagrangian
\begin{eqnarray}
{\cal L} &=& -\frac{1}{2g}\left\{(\partial\Lambda_3)^2 - (\partial\Phi_3)^2\right\}-\frac{1}{2g'}\left\{(\partial\lambda)^2 - (\partial\phi)^2\right\}+\frac{1}{4\pi}(\partial\Phi_3)^2 +\frac{1}{2\pi}(\partial\phi)^2\nonumber\\
 && -\frac{1}{2}(\partial\theta_1)^2 -\frac{1}{2}(\partial\theta_2)^2 + g\mu^2\cos\left\{\sqrt{4\pi}(\theta_1+\theta_2)+2\Phi_3\right\},\label{effectiveL}
\end{eqnarray}
where $\mu$ is an infrared cutoff parameter with mass
dimension.\cite{Justin,J_P}  We put
\begin{equation}
 \Theta = \sqrt{4\pi}(\theta_1+\theta_2)+2\Phi_3,
\end{equation}
to change $\Phi_3\rightarrow \Theta$ and integrate over $\theta_i,
\lambda, \phi$  and $\Lambda_3$ to obtain
\begin{equation}
 {\cal L} = -\frac{1}{16\pi}\left(1+\frac{g}{2\pi}\right)(\partial\Theta)^2+g\mu^2\cos\Theta.  
\label{boson_lag}
\end{equation}
As mentioned before the ghost field $\phi$ is decoupled from other fields
and $\Phi_3$ is absorbed into $\Theta$, and there is no ghost field in
the above expression.

The simple expression (\ref{boson_lag}), however,  obviously disagrees with our previous
consequence in Ref.\citen{J_M}. The decomposition of the Lagrangian into Eqs.(\ref{L_0})
and (\ref{L_I}) violates $SU(2)$ symmetry and the fermion Jacobian
(\ref{Jacobian1}) is not correctly derived. To see this we introduce
auxiliary vector fields $A_{\mu}^a$ and $a_\mu$ to rewrite
(\ref{ini_L}) as
\begin{eqnarray}
 {\cal L}&=& \overline{\psi}\dslash\psi + j_{\mu}^aA_a^{\mu}-\frac{1}{2g}A_{\mu}^a A_a^{\mu}+ j_\mu a^\mu-\frac{1}{2g'}a^{\mu}a_{\mu}\nonumber\\ 
&=& \overline{\psi}\{\dslash + i\tau_a(\dslash\Lambda^a+\gamma^\mu\epsilon_{\mu\nu}\partial^\nu\Phi^a)+i(\dslash\lambda +\gamma^\mu\epsilon_{\mu\nu}\partial^\nu\phi)\}\psi  \nonumber\\
&& -\frac{1}{2g}\{\partial^\mu\Lambda_a\partial_\mu\Lambda_a-\partial^\mu\Phi_a\partial_\mu\Phi_a\}-\frac{1}{2g'}\{\partial^\mu\lambda\partial_\mu\lambda-\partial^\mu\phi\partial_\mu\phi\},\label{L_2}
\end{eqnarray}  
where we put
\begin{subeqnarray}
 A_\mu^a &=& \partial_\mu\Lambda_a +\epsilon_{\mu\nu}\partial^\nu\Phi_a,\\
a_\mu &=& \partial_\mu\lambda+\epsilon_{\mu\nu}\partial^\nu\phi.
\end{subeqnarray}
By the decoupling transformation (\ref{decouple_trans}), Lagrangian (\ref{L_2})
is transformed into
\begin{eqnarray}
 {\cal L}&=& \overline{\psi '}\{\dslash+i\sum_{i=1,2}\overslash{A}_i'\tau_i\}\psi'\nonumber\\
&&-\frac{1}{2g}\{\partial^\mu\Lambda_a\partial_\mu\Lambda_a-\partial^\mu\Phi_a\partial_\mu\Phi_a\}-\frac{1}{2g'}\{\partial^\mu\lambda\partial_\mu\lambda-\partial^\mu\phi\partial_\mu\phi\},
\end{eqnarray}
where we put
\begin{eqnarray}
\sum_{i=1,2} \overslash{A}'_i\tau_i &\equiv& \sum_{i=1,2}e^{i\tau_3(\Lambda_3+\gamma_5\Phi_3)}\overslash{A}_i\tau_i e^{-i\tau_3(\Lambda_3-\gamma_5\Phi_3)}\nonumber\\
&=& \sum_{i=1,2}e^{2i\tau_3(\Lambda_3+\gamma_5\Phi_3)}\overslash{A}_i\tau_i.
\end{eqnarray}
To derive the Jacobian for the decoupling transformation
(\ref{decouple_trans})  we use the eigenfunction $\psi_n(x;t)$ which satisfies
\begin{equation}
 i\overslash{D}'(t)\psi_n (x;t)\equiv i(\dslash + i\overslash{\mbold{A}}(t))\psi_n(x;t)=\lambda_n\psi_n(x;t),\label{eigen_eq2}
\end{equation}
instead of (\ref{eigen_eq1}), where we define $\overslash{\mbold{A}}(t)$ as
\begin{eqnarray}
 \overslash{\mbold{A}}(t)&=& U(-\Lambda_3,\Phi_3,-\lambda,\phi;-t)\overslash{\mbold{A}}U(\Lambda_3,\Phi_3,\lambda,\phi;-t)\nonumber\\
&&-iU(-\Lambda_3,\Phi_3,-\lambda,\phi;-t)\dslash U(\Lambda_3,\Phi_3,\lambda,\phi;-t), \label{A(t)}
\end{eqnarray}
with
\begin{equation}
 \overslash{\mbold{A}}= \sum_{a=1}^3\overslash{A}_a\tau_a+\overslash{a}.
\end{equation}
Eq.(\ref{A(t)}) is rewritten as
\begin{equation}
 \overslash{\mbold{A}}(t) = (1-t)(\overslash{A}_3\tau_3 + \overslash{a})+ \exp \{2it\tau_3(\Lambda_3 + \gamma_5\Phi_3)\}\sum_{i=1,2}\overslash{A}_i\tau_i.
\end{equation}
We see $\overslash{\mbold{A}}(1)=\sum_{i=1,2}\overslash{A}'_i\tau_i$. The equation (\ref{eigen_eq2}) corresponds to the Euler equation for the
fermion field obtained from the Lagrangian (\ref{L_2});
\begin{equation}
 i(\dslash+i\overslash{\mbold{A}})\psi=0.
\end{equation} 

The fermion measure is transformed under (\ref{decouple_trans}) as
\begin{eqnarray}
 d\psi&=&|\det U(\Lambda_3,\Phi_3,\lambda,\phi)|d\psi',\label{det_a}\\
d\overline{\psi} &=& |\det U(-\Lambda_3,\Phi_3,-\lambda,\phi)|d\overline{\psi'}.\label{det_b}
\end{eqnarray}
The determinant of the above expression (\ref{det_a}) is derived as\cite{J_M}
\begin{equation}
 |\det U(\Lambda_3,\Phi_3,\lambda,\phi)|=\exp\frac{i}{4\pi}\int d^2x W,
\end{equation}
with
\begin{eqnarray}
W &=& \int_0^1 dt{\rm Tr}\left[\gamma_5(\phi+\tau_3\Phi_3)\{\dslash\overslash{\mbold{A}}(t)+ i\overslash{\mbold{A}}(t)\overslash{\mbold{A}}(t)\}\right]\nonumber\\
&=& \int_0^1 dt {\rm Tr}[\gamma_5(\phi+\tau_3\Phi_3)\{-(1-t)\epsilon_{\mu\nu}\gamma_5(\partial^\mu A_3^\nu\tau_3+\partial^\mu a^\nu) \nonumber\\
&& +\tau_3\gamma_5\epsilon_{ij}\epsilon_{\mu\nu}A_i^\mu A_j^\nu \cos (2t\Phi_3)-\tau_3\gamma_5A_i^\nu A_j^\nu\delta_{ij}g_{\mu\nu}\sin (2t\Phi_3)\}] \nonumber\\
&=& -\left[\frac{1}{2}\epsilon_{\mu\nu}\partial^\mu A_3^\nu\Phi_3+\epsilon_{\mu\nu}\partial^\mu a^\nu\phi\right. \nonumber\\
&&\left. -\frac{1}{2}\epsilon_{ij}\epsilon_{\mu\nu}A_i^\mu A_j^\nu\sin (2\Phi_3)+\frac{1}{2}g_{\mu\nu}A_i^\mu A_i^\nu\{1-\cos (2\Phi_3)\}\right]\nonumber\\
&=& \left[\frac{1}{2}(\partial\Phi_3)^2+(\partial\phi)^2 + \epsilon_{ij}\partial^\mu\Lambda_i\partial_\mu\Phi_j\sin (2\Phi_3)\right.\nonumber\\
&& \left. -\frac{1}{2}(\partial^\mu\Lambda_i\partial_\mu\Lambda_i-\partial^\mu\Phi_i\partial_\mu\Phi_i)\{1-\cos (2\Phi_3)\}\right].
\end{eqnarray}
The determinant (\ref{det_b}) is derived samely. Then the effective
Lagrangian including these determinants is given by
\begin{eqnarray}
{\cal L} &=& \overline{\psi'}(\dslash + i\overslash{A}_i'\tau_i)\psi' -\frac{1}{2g}\left\{(\partial\Lambda_a)^2 - (\partial\Phi_a)^2\right\}-\frac{1}{2g'}\left\{(\partial\lambda)^2-(\partial\phi)^2\right\}\nonumber\\
&& +\frac{1}{4\pi} \left[(\partial\Phi_3)^2 + 2(\partial\phi)^2 \right.\nonumber\\
&& + 2\epsilon_{ij}\partial^\mu\Lambda_i\partial_\mu\Phi_j\sin (2\Phi_3)-(\partial^\mu\Lambda_i\partial_\mu\Lambda_i-\partial^\mu\Phi_i\partial_\mu\Phi_i)\{1-\cos(2\Phi_3)\}\bigr].\nonumber\\
&&
\label{correct_lag}
\end{eqnarray}
We can obtain the same effective Lagrangian as (\ref{effectiveL}) by the
path integration over $\Lambda_i$ and $\Phi_i$ if the last line in the
above equation does not exist.

\setcounter{equation}{0}
\section{Concluding remark} We discuss bosonization of the extended
Thirring model with $SU(2)$ symmetry. As in the textbook of
Ref.\citen{Justin}, we divide the current-current interaction
$g\sum_{a=1}^3 j^\mu_a j_\mu^a + g'j^\mu j_\mu$ into $gj^\mu_3j_\mu^3+
g'j^\mu j_\mu$ and $g\sum_{a=1}^2j^\mu_aj_\mu^a$. We convert the former
interaction terms into $j_\mu^3A^\mu_3 + j_\mu a^\mu$ with introducing
the auxiliary boson fields $A_\mu^3$ and $ a_\mu$ and eliminate them
from the action by the decoupling transformation. Calculating the
Jacobian induced by the decoupling transformation and perturbative
expansion with respect to the latter interaction terms, we obtain the
corresponding boson model (\ref{boson_lag}). We note, however, that such
derivation of the Jacobian for the decoupling transformation violates
the $SU(2)$ symmetry, and we derive the correct expression
(\ref{correct_lag}) for the effective Lagrangian. If we eliminate
$\overline{\psi'}i\overslash{A}_i'\tau_i\psi'$ in (\ref{correct_lag}) by
a further decoupling transformation, we will obtain the same Lagrangian
as in Ref.\citen{J_M}, although it has a much complicated form and seems
to contain ghost fields. Optimistically these fields may be absorbed to
the other fields and may finally disappear from the effective
Lagrangian as in the ordinary Thirring model, which is open for future investigation.

\end{document}